\documentclass[printer]{aamod}

\usepackage{graphicx}
\usepackage[varg]{txfonts}
\usepackage[colorlinks]{hyperref}           
\usepackage{natbib} 
\bibpunct{(}{)}{;}{a}{}{,} % to follow the A&A style
\usepackage{lscape}

\begin{document}

\title{The \ion{Mg}{i} b triplet and the 4571 {\AA} line as diagnostics of
stellar chromospheric activity}

\author{C. Sasso \and V. Andretta \and L. Terranegra \and M. T. Gomez}

\offprints{C. Sasso, \email{csasso@oacn.inaf.it}}

\institute{INAF-Osservatorio Astronomico di Capodimonte, Salita Moiariello 16,
  I-80131 Napoli, Italy}

\date{Received / Accepted}

\abstract{The \ion{Mg}{i} 4571~{\AA} line and the $b$ triplet are denoted in the literature as diagnostics of
  solar and stellar activity since their formation is in the low
  chromosphere.}{To investigate the potential of these four spectral
  lines as diagnostics of chromospheric activity in solar-like stars, studying
  the dependence of the intensity of these lines from local atmospheric
  changes by varying atmospheric models and stellar parameters.}{Starting with Next-Gen photospheric models, we build a grid of atmospheric models
  including photosphere, chromosphere and transition region and solve the
  radiative transfer to obtain synthetic profiles to compare with observed
  spectra of main-sequence, solar like stars with effective temperatures in
  the range $4800-6400$~K, solar gravity and solar metallicity.}{We find that
  the \ion{Mg}{i} 4571~{\AA} line is significantly sensitive to local
  changes in the atmospheric model around the minimum temperature. Instead, the
  lines of the $b$ triplet do not show significant responses to
  changes on the local atmospheric structure.}{}

\titlerunning{\ion{Mg}{i} b triplet and 4571~{\AA} line as diagnostics of
  chromospheric activity}

\maketitle

\section{Introduction}\label{sec:introduction}

Stellar chromospheres are classically defined as an atmospheric layer lying
above the photosphere and below the corona, characterized by a positive
temperature gradient and a marked departure from radiative
equilibrium. Stellar chromospheres exhibit a wide range of phenomena
collectively called ``activity'', mainly due to the presence and time
evolution of highly structured magnetic fields emerging from the convective
envelope. Strong spectral lines are commonly used as diagnostics of
chromospheric activity, such as \ion{Ca}{ii} H and K lines and the infrared
triplet, the H$\alpha$ line whose core is formed in the chromosphere, the
\ion{Na}{i} D lines, and many UV lines from atoms and ions such as \ion{O}{i},
\ion{C}{i}, and \ion{Fe}{ii}, with formation temperature at
$\sim$10$^4$~K. 

Historically, magnetic activity in stellar atmospheres has been
confirmed from long-term observations of the spectroscopic and photometric
behavior of Sun-like stars in the \ion{Ca}{ii} H and K lines. In particular,
most of the data have been collected in two projects, one at Mount Wilson
Observatory \citep[MWO;][]{wilson,duncan} and another at Lowell Observatory
\citep{hall}. The observed line emissions, resulting from the non-thermal
heating that occurs in the presence of strong magnetic fields, are quantified
through the $S$ \citep{wilson} and $R'_{HK}$ \citep{noyes} indices. 

In this work, we investigate the potential of four \ion{Mg}{i} spectral lines,
the 4571~{\AA} line and the $b$ triplet at 5183, 5172 and 5167~{\AA}, as
diagnostics of chromospheric activity in solar-like stars. These lines have
been modeled and observed in the solar atmosphere by many authors
\citep[i.e.,][]{athay,canfield,cannon,white,altrock,heasley} and are known as
spectroscopic probes of physical conditions in the solar upper photosphere and
lower chromosphere. For these reasons they are well suited as a diagnostic of the low chromosphere, close to the
temperature minimum region. \citet{langangen} studied the \ion{Mg}{i}
4571~{\AA} theoretical line formation in the quiet Sun and in a series of
sunspot models, concluding that the line can be used to probe the lower
chromosphere, especially for cooler atmospheres, such as sunspots. 

The \ion{Mg}{i} 4571~{\AA} line is a "forbidden", intersystem line originating between the
atomic levels $3^1S$ and $3^3P_1$ and, in the Sun, forms at
about 500~km above $\tau_{5000}=1$ (see also discussion in Sec.~\ref{sec:atomicmodel} and in particular the lower panel of Fig.~\ref{fig:opacity}). Its source function is very
close to local thermodynamical equilibrium (LTE) because the population \emph{ratio} of its upper and lower levels is determined mainly by collisional processes. Despite its small oscillator strength ($2.4\times 10^{-6}$) it is a prominent
line in the solar spectrum because its opacity is determined by the population of the \ion{Mg}{i} ground state. This also implies that non local thermodynamical equilibrium (NLTE) effects may be important in determining its optical depth. Hence the need of full, NLTE, multi-level calculations. 

The \ion{Mg}{i} $b$ triplet originates between the atomic levels $3^3P_{0,1,2}$
and $4^3D$, and comprises a component at 5183~{\AA} with $J_l=2$ ($b_1$), one
at 5172~{\AA} with $J_l=1$ ($b_2$) and one at 5167~{\AA} with $J_l=0$
($b_3$).  The relative simplicity of the formation of the \ion{Mg}{i} 4571 line is in contrast with the complex formation of the \ion{Mg}{i} $b$ triplet. In the solar atmosphere the cores of the $b$
triplet lines form in the low chromosphere. Both their source function and opacities are affected by NLTE, mainly though photoionization. The coupling with other atmospheric regions introduced by this kind of NLTE effects makes these lines less sensitive to local conditions. The wings of these lines form in the photosphere \citep{zhao2,sasso}.
 
The behavior and the formation of these Mg lines have been studied also in
stellar atmospheres and, in particular, the \ion{Mg}{i} $b$ triplet has been often
indicated in the literature as a good diagnostic of stellar activity
\citep{monteslibrary,montes,basri}. Recently, \citet{osorio1} and \citet{osorio2} studied
the Mg line formation in late-type stellar atmospheres, using the same
numerical code to solve the coupled equations of radiative transfer and
statistical equilibrium we are using in this work and describe in
Sec.~\ref{sec:model}. They constructed a new Mg model atom for NLTE studies
and implemented new quantum mechanical calculations to investigate the effect
of different collisional processes in benchmark late-type stars. Their study
aims to give abundance corrections especially in metal-poor stars, where
departures from LTE affect line formations. A similar attempt has been done by
\citet{merle1} for the Mg line formation in late-type giant and supergiant
stars, and in particular for the lines we are studying, also in a metal-poor
dwarf and a metal-poor giant \citep{merle2,merle3}.

In this work, we extend the above investigation to solar-like stars (dwarf stars of
spectral types F-K and solar metallicity) by studying the
dependence of the intensity of these lines from local atmospheric changes by
varying atmospheric models and stellar parameters. The aim is to evaluate the
reliability of these four Mg spectral lines as diagnostics of chromospheric
activity in this type of stars. 

The paper is structured as follows. In Sec.~\ref{sec:model} we describe the Mg
atomic model and the stellar atmospheric models we used in this work. In
Sec.~\ref{sec:results}, we compare the synthetic profiles obtained with eight
observed spectra of Sun-like stars and discuss the sensitivity of the Mg lines
to local changes in the atmospheric model. Finally, in the conclusions
(Sec.~\ref{sec:conclusions}), the role of these Mg lines as diagnostic of
chromospheric activity is defined.
%vedere se bisogna linkare gli oggetti stellari a simbad (aadoc.pdf)

\section{Model calculations and solar test}\label{sec:model}

In order to investigate the sensitivity of the \ion{Mg}{i} 4571~{\AA} and the
$b$ triplet line profiles in main-sequence, solar-like stars to local
atmospheric changes at chromospheric levels, we built a grid of models of a
stellar atmosphere with different effective temperatures, solar gravity and
solar metallicity and solved the coupled equations of radiative transfer and
statistical equilibrium were solved for the \ion{Mg}{i} using version 2.2
of the code {\small{MULTI}} \citep{multi}. 

\subsection{Atomic model}\label{sec:atomicmodel}

In this work we used, with some modifications, the \ion{Mg}{i} atomic model
proposed by \citet{carlsson}, consisting of 66 levels (including the
continuum), 315 lines and 65 bound-free ($b-f$) transitions treated in
detail. The atomic model was slightly modified by Carlsson (private
communication) with respect to the original one presented in the above publication. The modifications were done in order to analyze in detail the Mg $b$
triplet that it is treated as a superposition of three lines, taking into
account that they are close enough to affect each other. The photo-ionization
cross-sections are given with many more wavelength points which makes the
line blanketing calculation more accurate. In the literature we found other Mg
atomic models containing different numbers of atomic levels. \citet{Mg} used a
12-level atomic model to study the influence that the uncertainties in the
value of different atomic parameters may have on the calculated profiles,
comparing them with the Sun spectrum. \citet{zhao2} and \citet{zhao1}
presented an 84-level atomic model to investigate the formation of neutral Mg
lines in the solar photosphere and in the photosphere of cool stars,
respectively. Their model is nearly the same as that used by
\citet{carlsson}. \citet{mashonkina} and \citet{shimanskaya} instead, used a
49-level \ion{Mg}{i} atomic model to study the formation of Mg lines in the
atmospheres of stars of various spectral types performing a detailed
statistical-equilibrium analysis. Finally, there are two recent atomic models
constructed by \citet{merle1} and \citet{osorio1}, used to investigate the
effect of NLTE Mg line formation in late-type stellar atmospheres, as
described in Sec.~\ref{sec:introduction}. Results from these papers show that
departures from LTE for the lines we are considering are significant for low
metallicity stars. We decided to use the model proposed by \citet{carlsson}
because its original use is closer to our study and treats specifically the Mg
$b$ triplet. Moreover, we want to compare the results of the line synthesis
with observed spectra from solar-like stars (stars with solar metallicity). All the
atomic parameters used are specified in the work of \citet{carlsson}. 

In order to match the observed profiles and test the procedure on the Sun, we
recalculated the Van der Waals damping parameter for the transitions involved
in the formation of the lines of our interest. As pointed out by \citet{Mg},
the Van der Waals broadening dominates over the radiative and the Stark
broadenings in determining the Voigt coefficient in the formation of the
5173~{\AA} line (studied as representative of the whole triplet). We
recalculated also the Van der Waals damping parameter for the 4571~{\AA} line,
for homogeneity. The value of the theoretical Van der Waals damping
parameter, $\Gamma_{\rm VW}$, has been calculated with two different methods,
following the theoretical models presented by \citet{vdw1} and by
\citet{vdw2}. The values retrieved are respectively 2.55 and 4.43
($10^{-8}$~rad~cm$^3/$s) for the $b$ triplet and 0.86 and 1.25
($10^{-8}$~rad~cm$^3/$s) for the 4571~{\AA} line.
The Van der Waals damping coefficient required in the
atomic model is the factor $f_{\rm GVW}$ in the following formula
\begin{eqnarray}
\Gamma_{\rm VW}=f_{\rm GVW}\;\Gamma_{\rm VWM},
\end{eqnarray}
where $\Gamma_{\rm VWM}$ is the value of the broadening calculated by the program 
MULTI, as follows \textbf{\citep[][Sec.~9.3]{mihalas}}
\begin{eqnarray}
\Gamma_{\rm VWM}=8.08\left(1+0.41\left(\frac{N_{He}}{N_H}\right)\right)V_H^{0.6}\;N_H\;C_6^{0.4},
\end{eqnarray}
where
\begin{eqnarray}
V_H^{0.6}=\left[\frac{8kT}{\pi}\left(\frac{1}{m_H}+\frac{1}{m_{Mg}}\right)\right]^{0.3}
\end{eqnarray}
and
\begin{eqnarray}
C_6=1.01\;10^{-32}(13.6\;I)^2\left[\frac{1}{(E_c-E_u)^2}-\frac{1}{(E_c-E_l)^2}\right].
\end{eqnarray}
The quantities $N_{He}$ and $N_H$ are, respectively, the helium and hydrogen
atomic number density ($N_{He}/N_H=0.1$), $k$ is the Boltzmann constant, $T$
is the temperature in K, $m_H$ and $m_{Mg}$ are, respectively, the atomic
weight of hydrogen and magnesium in g, $I$ is the ionization state of the
lower transition level, $E_c$, $E_u$ and $E_l$ are the continuum, upper and
lower level energy.

We adopt the values of the Van der Waals damping ($f_{\rm GVW}=\Gamma_{\rm
  VW}/\Gamma_{\rm VWM}$) that give the better agreement with the solar observed profiles: $f_{\rm GVW}=2.3$ for the $b$ triplet
\citep[the same value used by][]{carlsson} and $f_{\rm GVW}=1.1$ for the 4571~{\AA} line (in
the original atomic model was 10.9).

As it is pointed out in \citet{langangen} and \citet{carlsson}, it is 
important to treat the line blanketing in order to get the ionization balance 
right. For this reason, some treatment of line-opacity needs to be included in
the calculations. The opacity package that comes with the code takes into
account free-free opacity, Rayleigh scattering, and bound-free transitions
from hydrogen and metals. The absence of line blanketing in the standard
opacity package leads to an overestimate of emerging intensities (especially
in the UV) and therefore to an overestimate of photoionization rates, some of
which can be important in NLTE calculations. We have estimated the effect of
line blanketing by increasing the background standard opacity by a suitable
factor, following the method described in \citet{busa}.
This method is based on a semi-empirical approach: any background opacity source missing from the calculations is estimated from the ratio of the calculated flux to the observed one \citep[in this work, we take the Next-Gen fluxes][as our "observed" reference spectrum]{nextgen}. This approach therefore provides a quick, albeit approximate way to take into account line haze, either in LTE or NLTE, in the most relevant photoionization rates. In particular, there is no guarantee that the depth dependence of the photoionization rates is correctly computed throughout the atmosphere. But this method already provides some improvements on the modeling of the \ion{Mg}{i} lines as we show below.

\begin{figure*}
\centering
\includegraphics[clip=true,width=12cm]{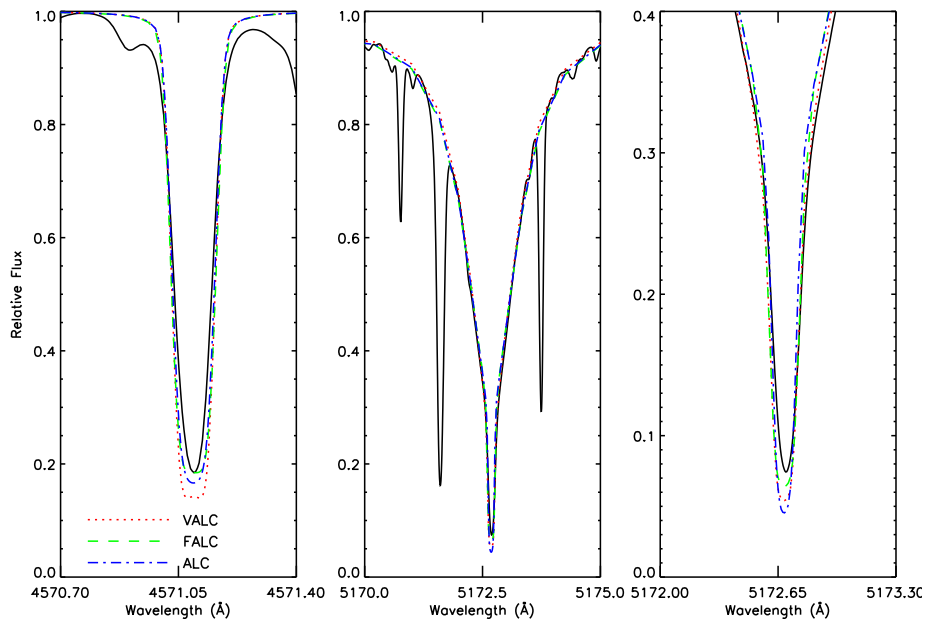}
\caption{\ion{Mg}{i} 4571~{\AA} (left panel) and 5173~{\AA} ($b_2$, middle and
  right panel) observed and synthetic line profiles for the Sun. The
  third panel magnifies the $b_2$ core. The black solid lines are the
  observed profiles \citep[Kitt Peak solar flux atlas,][]{atlas} while
  the red dotted, green dashed and blue dash-dotted lines are the synthetic
  profiles obtained by using three different atmospheric models, VALC, FALC
  and ALC, respectively.}
\label{fig:righe_sun}
\end{figure*}

We tested our procedure and data set on the Sun using three different 
quiet-Sun atmospheric models. The three semi-empirical models of the solar 
atmosphere used, are the ones proposed by \citet[VALC,][]{VAL3C}, 
\citet[FALC,][]{falc}, and \citet[ALC,][]{alc7}, encompassing both the photosphere and the
chromosphere. Figure~\ref{fig:righe_sun} shows the synthetic \ion{Mg}{i}
4571~{\AA} (left panel) and $b_2$ (middle and right panel) line profiles obtained using
the \ion{Mg}{i} atomic model and the three atmospheric models (VALC red dotted,
FALC green dashed, and ALC blue dash-dotted line), compared with the Kitt Peak solar flux
atlas \citep[black solid line,][]{atlas}. The synthetic profiles are broadened for
the value of the projected rotational velocity, $v \sin i= 1.7$~km/s
\citep{liu}. The differences in the atmospheric models give a first indication
of the sensitivity of the studied lines to changes in the atmosphere. The VALC model has the lowest minimum temperature value and the steepest temperature gradient towards the minimum. Therefore, the 4571~{\AA} line, forming in LTE around the minimum temperature region, shows a deeper core. Indeed, the main correction made to the VALC model that brought to the construction of new models, like the FALC and the ALC, was the inclusion of more line haze lines so that the temperature gradient towards the minimum became less steep from less overionization. The $b$ triplet does not seem to be affected by any change in the atmospheric model. The best agreement with the observed spectrum is obtained with the FALC model that has a higher minimum temperature value and a less steep temperature gradient towards the minimum with respect to the VALC model.

\begin{figure}[htbp]
\centering
\includegraphics[clip=true,width=8cm]{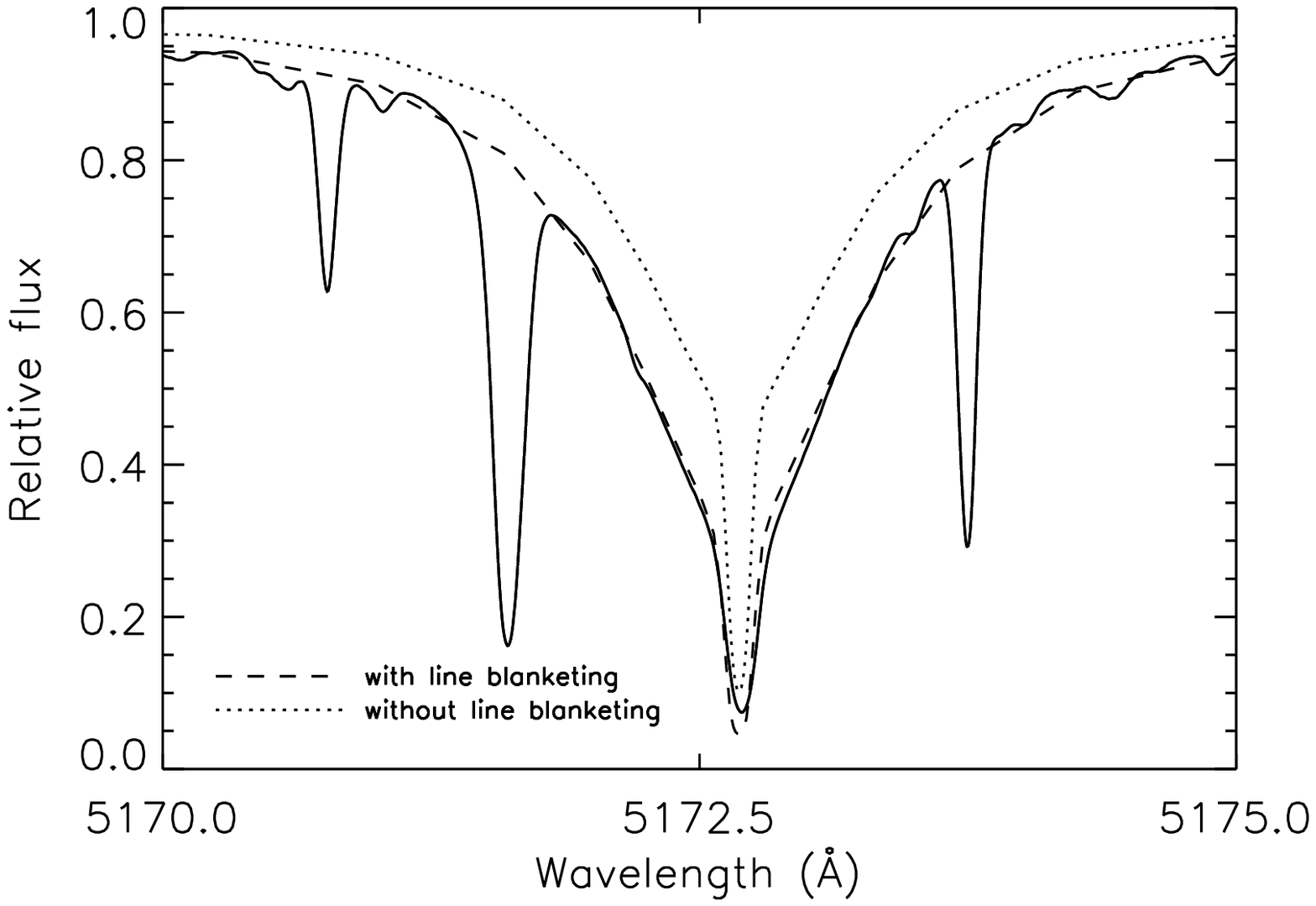}
\includegraphics[clip=true,width=8cm]{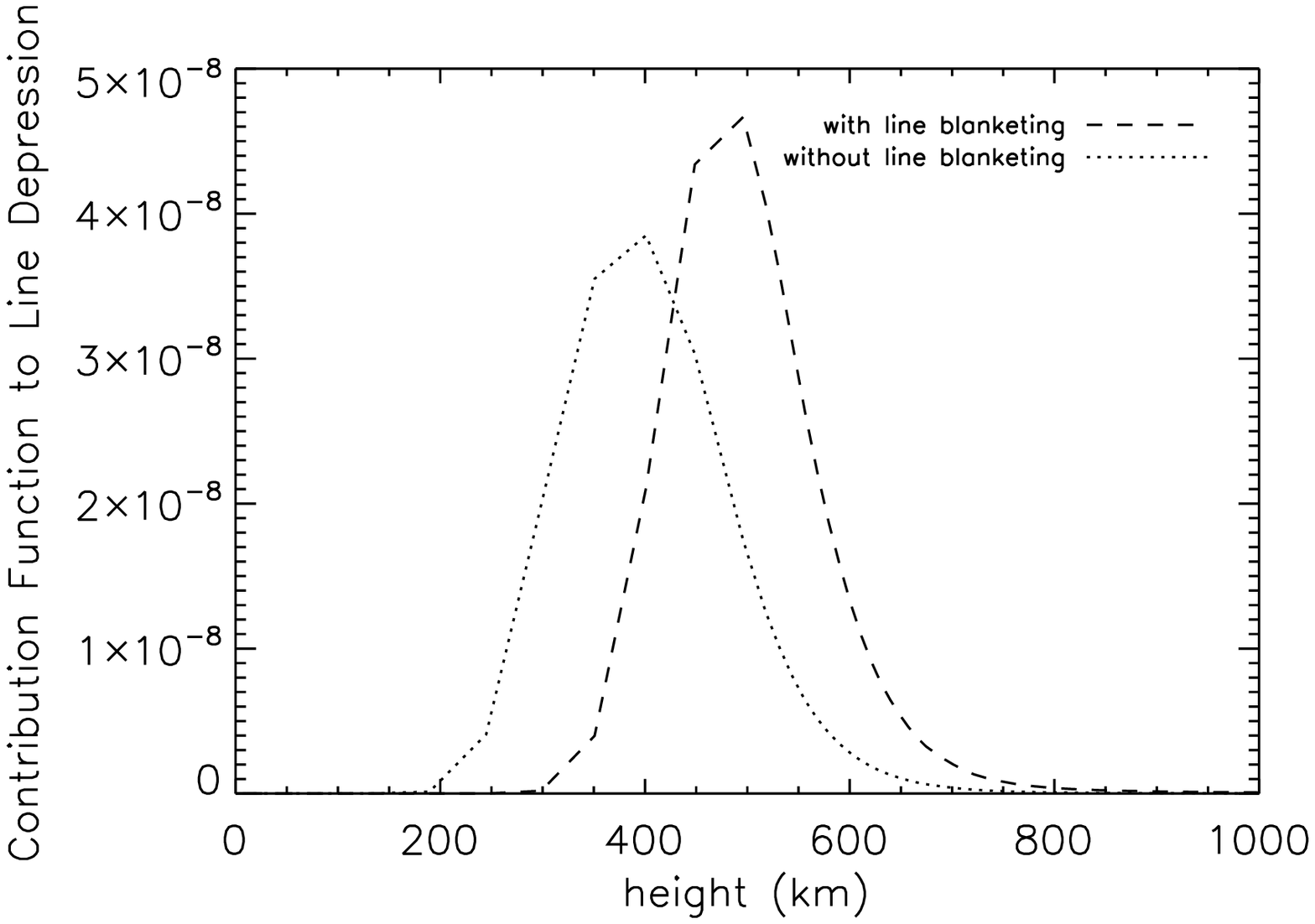}
\caption{Upper panel: synthetic profiles of the \ion{Mg}{i} $b_2$ line obtained with the ALC model of the solar atmosphere with (dashed line) and without (dotted line) line blanketing. The solid line is the observed profile. Lower panel: contribution function to line-depression (erg~cm$^{3}$~s$^{-1}$~{\AA}$^{-1}$) for the 4571~{\AA} line versus height, for the same atmospheric model as in the upper panel, with (dashed line) and without (dotted line) line blanketing.}
\label{fig:opacity}
\end{figure}

We also made some tests to show how the method to treat the background line-opacity, described in \citet{busa}, improves the modeling of the \ion{Mg}{i} lines. In Fig.~\ref{fig:opacity} (upper panel) we plot, as an example, the synthetic profiles of the $b_2$ line obtained with the ALC model of the solar atmosphere with (dashed line) and without (dotted line) line blanketing. The solid line is the observed profile. For the 4571~{\AA} line, in the lower panel of Fig.~\ref{fig:opacity}, we plot the contribution function to line depression \citep{magain} versus height, for the same atmospheric model as in the upper panel with (dashed line) and without (dotted line) line blanketing. When we do not take into account any correction to the {\small{MULTI}} background opacity, the line forms at lower heights in the solar atmosphere, and therefore becomes less sensitive to changes in the chromosphere and at the temperature minimum (see discussion in Sec.~\ref{sec:results}).

\subsection{Atmospheric models}\label{sec:atmosphere}

\begin{figure}
\centering
\includegraphics[clip=true,width=8cm]{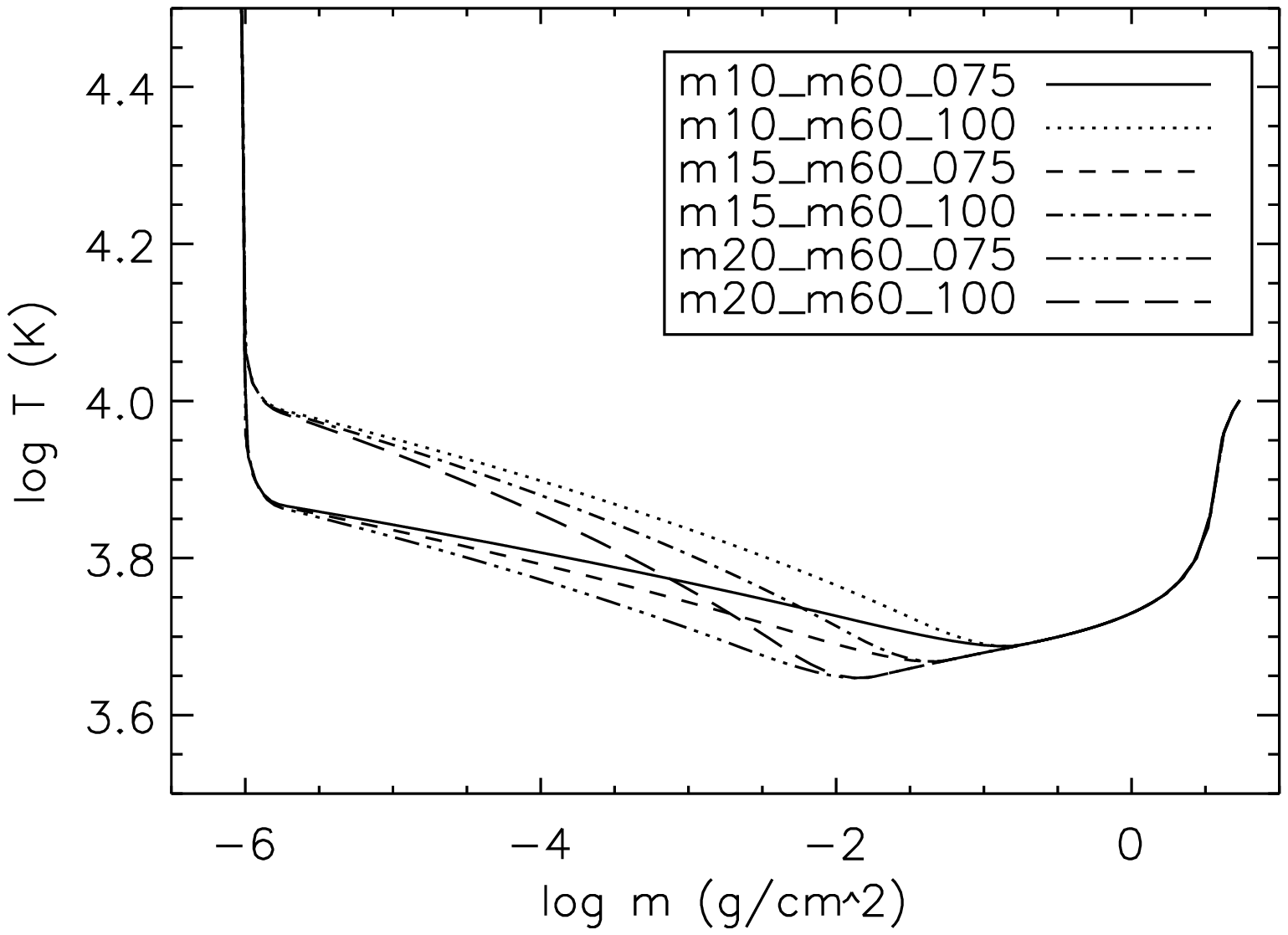}
\caption{Atmospheric models obtained by adding a chromosphere and a transition 
region to the Next-Gen photosphere with $T_{\rm{eff}}=6200$~K. The six models 
differ in the chromospheric temperature gradient and/or the temperature 
minimum value as indicated in Table~\ref{tab:valori_atm}.} 
\label{fig:atm_models}
\end{figure}

We selected from the Next-Gen database a grid of photospheric
models of a stellar atmosphere with effective temperatures
$T_{\rm{eff}}= 4800, 5200, 5600, 5800, 6200$ and $6400$~K, solar gravity
and solar metallicity, i.e. $\log g=4.5$ and
$\left[\textmd{A}/\textmd{H}\right]=0.0$. Starting from the photospheric
models, we built a grid of atmospheres including photosphere, chromosphere and
transition region (TR). The chromosphere is represented by a linear dependence
of the temperature as a function of $\log m$. We indicate with $\log m_0$
(column mass value) the point where it starts in the photosphere and with the
coordinates ($\log m_1$, $T_1$) the point where the chromosphere ends and the
TR starts. By giving the value of $\log m_0$, we fix the minimum value of the
temperature in the atmospheric model, and, by giving ($\log m_1$, $T_1$) we
fix the thickness of the chromosphere and the temperature gradient. We also
need to specify the temperature gradient in the TR by giving the value of the
last point of the model ($\log m_2$, $T_2$). We used, as for the chromospheric
segment, a linear dependence of $T$ as a function of $\log m$ and for all the
models the transition region gradient is the same. The values of $\log m_0$,
$\log m_1$, $T_1$, $\log m_2$ and $T_2$ are indicated in
Tab.~\ref{tab:valori_atm} for the six models created from the same starting
photosphere for each photospheric model, identified by the effective
temperature, $T_{\rm{eff}}$. We refer to the different models with
the names indicated in the first column, given by the values of $\log m_0$,
$\log m_1$, and $T_1$. The names of the models are defined as mxx\_myy\_zzz
where xx$=-\log m_0\times10$, yy$=-\log m_1\times10$ and zzz$=T_1/100$. The
most active models are the ones with a higher value of $\log m_0$ (the
chromosphere starts deeper in the photosphere) and/or a steeper chromospheric
temperature gradient. Fig.~\ref{fig:atm_models} shows as an example six
atmospheric models created by starting from a photosphere with $T_{\rm{eff}}=6200$~K, $\log g=4.5$ and
$\left[\textmd{A}/\textmd{H}\right]=0.0$. We did not consider the possibility
of $\log m_1$ greater than $-6$. Considering the formation depth in the solar
atmosphere of the lines studied in this paper, their profiles should not be
influenced by a change in the atmospheric model occurring in the upper
chromosphere.
\begin{table}
\caption{Values of $\log m$ (g/cm$^2$) and $T$ (K) used to build the six 
atmospheric models starting from the Next-Gen photospheres. $\log m_0$ is the 
starting point of the chromospheric segment in the photosphere (minimum 
temperature of the model). ($\log m_1$, $T_1$) are the coordinates of the end 
point of the chromospheric segment; the transition region ends at the point 
($\log m_2$, $T_2$).}
\label{tab:valori_atm} \centering
\begin{tabular}{cccccc}
\hline\hline
Model        & $\log m_0$ & $\log m_1$ & $\log m_2$ & $T_1$ & $T_2$  \\
\hline
m10\_m60\_075 &  -1.0    &  -6.0    &  -6.1    & 7500  & 100000 \\
m10\_m60\_100 &  -1.0    &  -6.0    &  -6.1    & 10000 & 100000 \\
m15\_m60\_075 &  -1.5    &  -6.0    &  -6.1    & 7500  & 100000 \\
m15\_m60\_100 &  -1.5    &  -6.0    &  -6.1    & 10000 & 100000 \\
m20\_m60\_075 &  -2.0    &  -6.0    &  -6.1    & 7500  & 100000 \\
m20\_m60\_100 &  -2.0    &  -6.0    &  -6.1    & 10000 & 100000 \\
\hline
\end{tabular}
\end{table}

\section{Results}\label{sec:results}

\subsection{Stellar spectra}

In order to compare the obtained synthetic profiles with observed ones, we
selected eight stellar spectra from the high resolution (R=80000) and high SNR
(300-500) UVES Paranal Observatory Project (UVES POP) library
\citep{bagnulo}. The chosen stars have solar gravity and solar metallicity and
effective temperatures in the range $\sim4800-6400$~K, show magnetic activity
and all the Mg lines under study are available in the UVES POP
library. Table~\ref{tab:stelle} presents the atmospheric parameters of the
selected stars collected from the literature. The first column shows the ID of
the star, the second column gives the effective temperature, the third column
provides the iron abundance and the fourth the surface gravity. The fifth
column gives the projected rotational velocity and the last column
gives the level of activity measured with two different activity index
$R'_{HK}$ or $R'_{8662}$, depending on the star and the measurement available
in the literature. Traditionally the \ion{Ca}{ii} H and K resonance lines
at 3968 and 3934 {\AA} have been used as chromospheric diagnostics. However,
observations of these lines in faint stars can be difficult, due to often low
stellar flux in the blue, and the relatively poor blue response of many CCD
detectors. In contrast, the \ion{Ca}{ii} infrared triplet lines at 8498, 8542,
and 8662 {\AA} are strong and their location in the red makes them more
suitable for CCD observations \citep{soderblom,andretta,busa1,marsdena}. The
values of gravity and metallicity are not available for all the stars in the
literature. (We recall that the atmospheric models are built starting from
photospheres of solar gravity and metallicity, $\log g=4.5$ and
$\left[\textmd{A}/\textmd{H}\right]=0.0$).
\begin{table*}
\caption{Temperature, gravity, iron abundance, projected rotational 
velocity and activity index. References:
$^{1}$\citet{soubiran},$^{2}$\citet{mermilliod}, $^{3}$\citet{dasilva},
$^{4}$\citet{marsdena}, $^{5}$\citet{dorazi}, $^{6}$\citet{battistini},
$^{7}$\citet{nordstrom}, $^{8}$\citet{henry},
$^{9}$\citet{casagrande}, $^{10}$\citet{stauffera},
$^{11}$\citet{marsdenb}, $^{12}$\citet{katsova}, $^{13}$\citet{cenarro}, $^{14}$\citet{glebocki}.} 
\label{tab:stelle} \centering
\begin{tabular}{ccccccc}
\hline\hline ID   & $T_{\rm{eff}}$ (K) & $\log g$ & 
$\left[\textmd{Fe}/\textmd{H}\right]$ & $v\sin i$ (km/s) & $\log\left(R'_{HK}\right)$ & $\log\left(R'_{8662}\right)$\\
\hline
HD 73722   & $6480^{1}$  & $4.30^{1}$  & $-0.05^{1}$ & $8.7\pm0.8^{2}$  & $-4.92^{3}$ & \\
IC 2391 73 & $6080^{4}$  & $4.45^{5}$  & $0.00^{5}$   & $8.4\pm0.7^{2}$ & & $-4.49^{4}$\\
HD 211415  & $5825^{6}$  & $4.40^{6}$  & $-0.23^{6}$  & $3^{7}$      &  $-4.86^{8}$ &\\
HD 74497   & $5642^{9}$  &            & $-0.12^{9}$  & $7^{7}$         &  $-4.81^{8}$ &\\
IC 2391 98 & $5201^{10}$ &            &             & $12.6^{2}$    & & $-4.89^{4}$\\
HD 149661  & $5216^{6}$  & $4.60^{6}$  & $-0.01^{6}$ & $2.2\pm0.4^{11}$ &$-4.57^{12}$ & \\
HD 22049   & $5135^{1}$  & $4.70^{1}$  & $-0.07^{1}$ &  $8^{13}$        & $-4.52^{12}$ &\\
HD 209100  & $4754^{1}$  & $4.45^{1}$  & $-0.20^{1}$  & $2.0^{14}$       &$-4.57^{8}$ &\\
\hline
\end{tabular}
\end{table*}

\begin{figure*}
\centering
\includegraphics[clip=true,width=12cm]{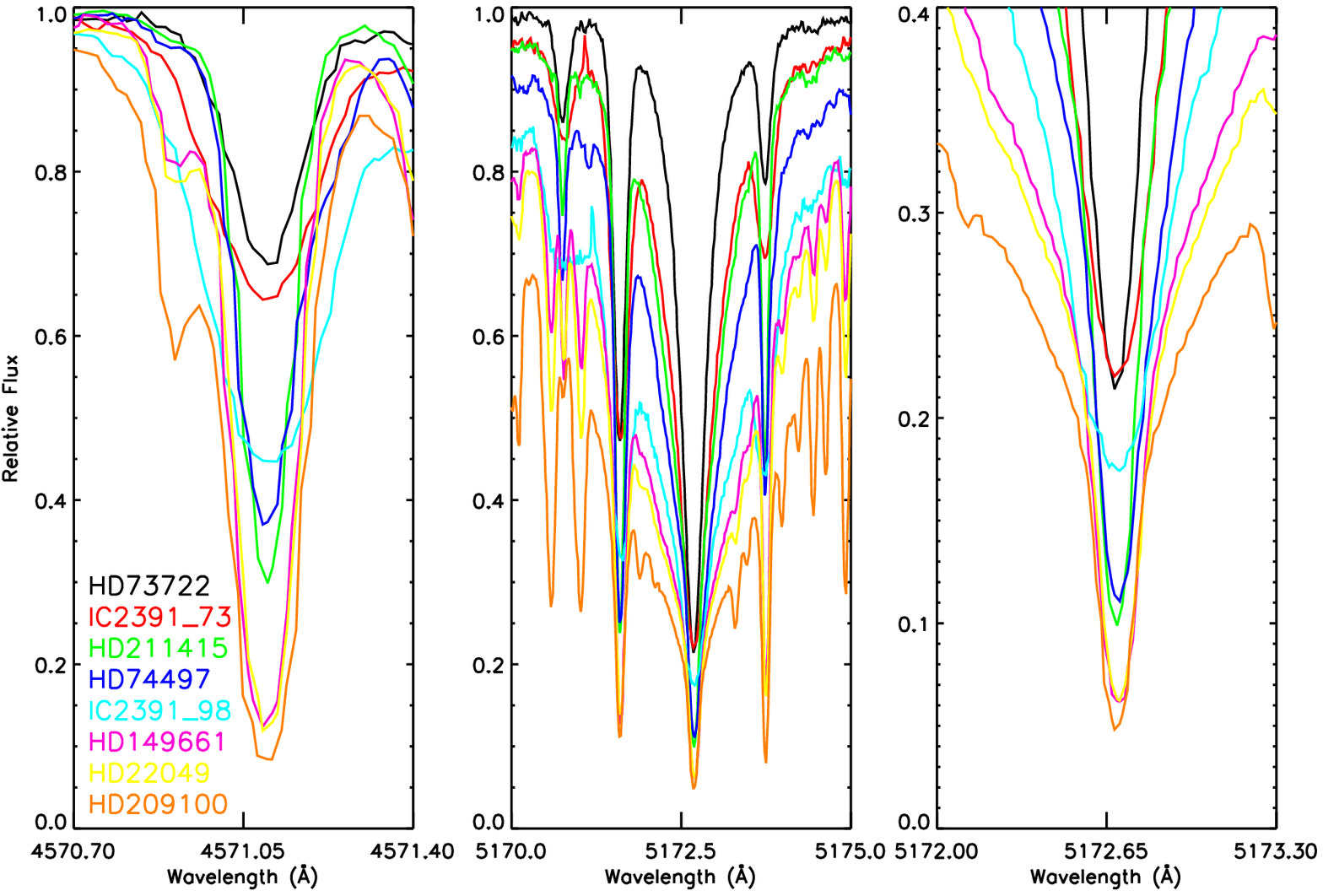}
\caption{\ion{Mg}{i} 4571~{\AA} (left panel), $b_2$ at 5173~{\AA} (middle) and
  $b_2$ core (right) observed line profiles for the stars in
  Tab.~\ref{tab:stelle}. The third panel magnifies the $b_2$ core.} 
\label{fig:stelle_obs}
\end{figure*}

Figure~\ref{fig:stelle_obs} shows the observed profiles of the stars in
Tab.~\ref{tab:stelle} for the spectral lines we are studying, the 4571~{\AA} line
(left panel) and the central line of the $b$ triplet at 5173~{\AA} (middle and
right panel). The profiles differ mainly because of the different effective
temperatures of the stars. Both lines show a less deep core and a larger width
with increasing temperature. The only star that displays a different behavior,
showing a less deep core than expected is the IC 2391 98. This happens because
the core depth and the width of the lines are influenced also by the value of
$v\sin i$ (less deep core and larger width with increasing projected
rotational velocity) and this star has the largest value of $v\sin i$ with
respect to all the other stars.

\subsection{Comparison with the stellar spectra}

\begin{figure}
\centering
\includegraphics[clip=true,width=9cm]{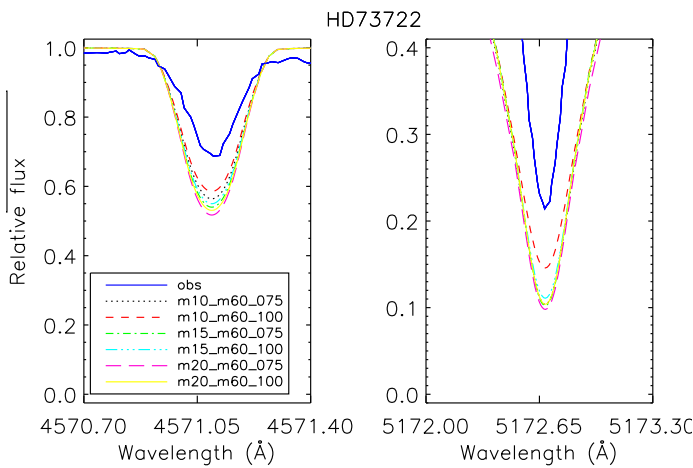}
\caption{\ion{Mg}{i} 4571~{\AA} (left panel) and 5173~{\AA} ($b_2$, right) 
observed and synthetic line profiles for the HD 73722 star. The solid thick 
lines are the observed profiles while the thin lines (with the same linestyle 
as in Fig.~\ref{fig:atm_models}) are the synthetic profiles obtained by using 
the different atmospheric models described in Table~\ref{tab:valori_atm}.}
\label{fig:star1}
\end{figure}

\begin{figure}[htbp]
\centering
\includegraphics[clip=true,width=9cm]{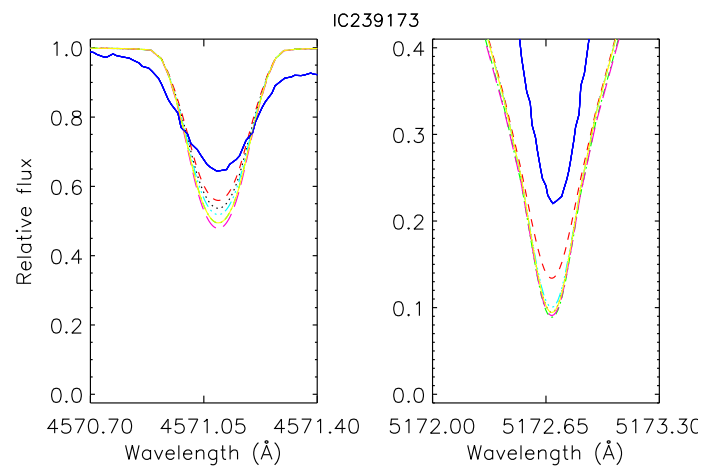}
\caption{As in Fig.~\ref{fig:star1} for IC 2391 73.}
\label{fig:star2}
\end{figure}

\begin{figure}[htbp]
\centering
\includegraphics[clip=true,width=9cm]{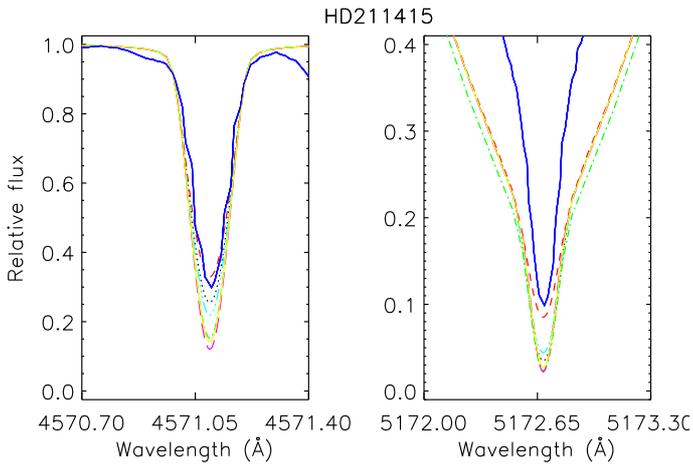}
\caption{As in Fig.~\ref{fig:star1} for HD 211415.}
\label{fig:star3}
\end{figure}

\begin{figure}
\centering
\includegraphics[clip=true,width=9cm]{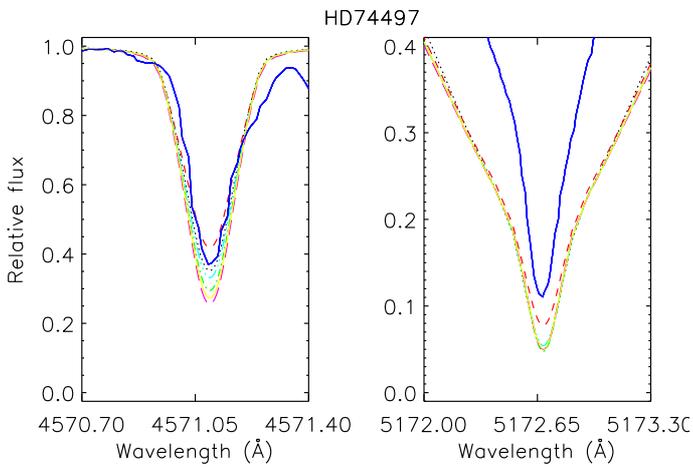}
\caption{As in Fig.~\ref{fig:star1} for HD 74497.}
\label{fig:star4}
\end{figure}

\begin{figure}[htbp]
\centering
\includegraphics[clip=true,width=9cm]{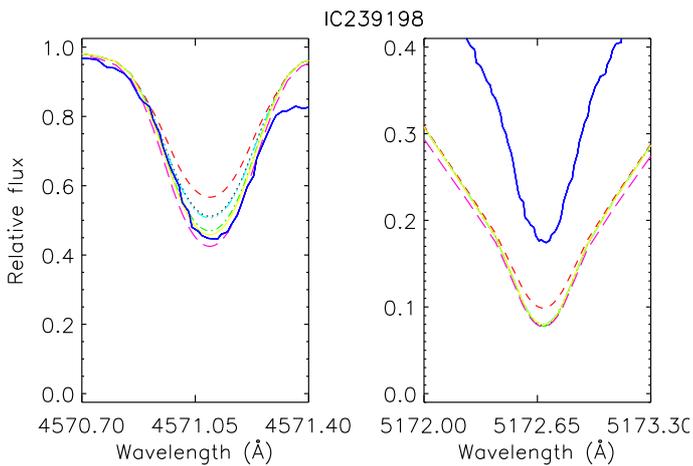}
\caption{As in Fig.~\ref{fig:star1} for IC 2391 98.}
\label{fig:star5}
\end{figure}

\begin{figure}[htbp]
\centering
\includegraphics[clip=true,width=9cm]{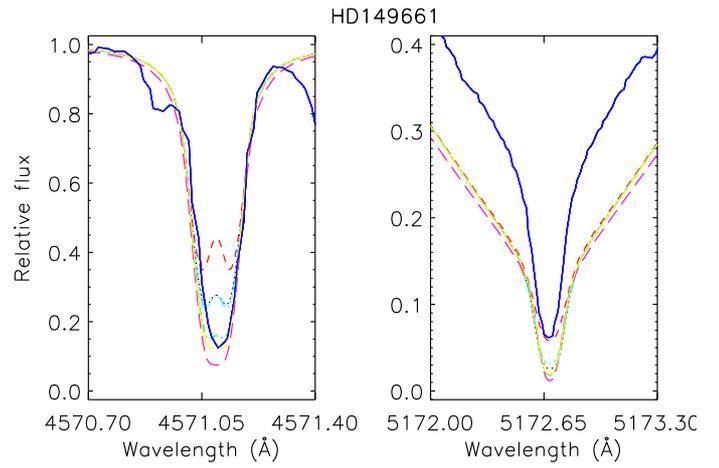}
\caption{As in Fig.~\ref{fig:star1} for HD 149661.}
\label{fig:star6}
\end{figure}

\begin{figure}[htbp]
\centering
\includegraphics[clip=true,width=9cm]{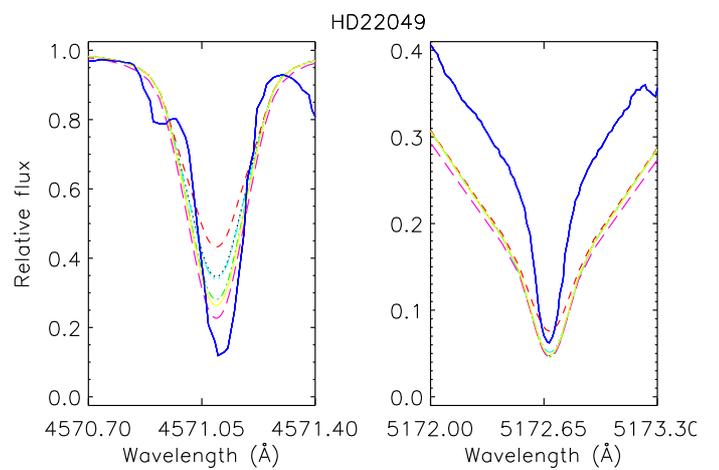}
\caption{As in Fig.~\ref{fig:star1} for HD 22049.}
\label{fig:star7}
\end{figure}

\begin{figure}[htbp]
\centering
\includegraphics[clip=true,width=9cm]{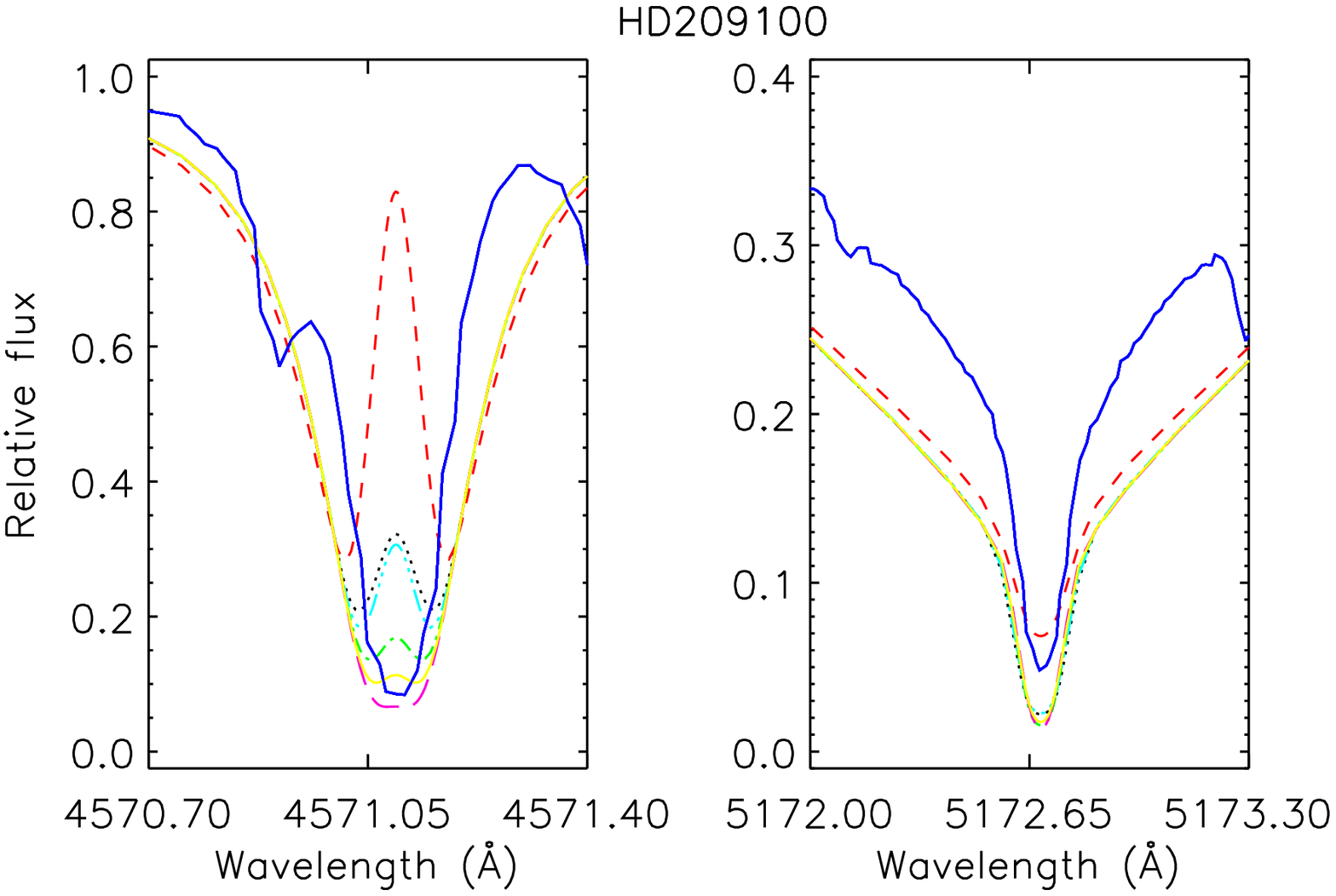}
\caption{As in Fig.~\ref{fig:star1} for HD 209100.}
\label{fig:star8}
\end{figure}

Comparison to the observed profiles are displayed in
Figs.~\ref{fig:star1}-\ref{fig:star8}. For each star we over-plot to the
observed profile (blue solid thick line), the synthetic profiles obtained with
the six different atmospheric models, built as described in
Sec.~\ref{sec:atmosphere} starting from the corresponding photospheric model
with the $T_{\rm{eff}}$ of the star. The synthetic profiles are plotted using the
same linestyles as in Fig.~\ref{fig:atm_models} and different colors to
distinguish between the atmospheric models. In Tab.~\ref{tab:parametri} we
indicate for each star, the value of $T_{\rm{eff}}$ used to create the atmospheric
models and the value of $v\sin i$ adopted, starting from the values reported
in the literature indicated in Tab.~\ref{tab:stelle}. In the last two columns
are indicated the atmospheric models that best fit the observed profiles of
the 4571~{\AA} and the $b_2$ line as deduced by a visual inspection of the
figure. All the synthetic profiles are broadened for the $v\sin i$ value
listed in Table~\ref{tab:parametri}. For the $b$ triplet we plot only the
$b_2$ line at 5172.68~{\AA}, since all three lines forming the triplet behave
in the same way. We show only the comparison with the core of the $b_2$ line, since we
are interested only in activity at chromospheric level where the core forms,
while the wings are photospheric. If we change the stellar parameters (i.e.,
pressure, gravity and/or metallicity) and/or the opacity routines as input to
the numerical code we could be able to define an ad-hoc photosphere for each
star but diagnosing the stellar parameters is outside the scope of this
work. Some of the observed profiles of the 4571~{\AA} line are blended in the blue wing with a \ion{Ti}{i} line. This line is identified in the solar spectrum at 4570.91~{\AA} \citep{moore} and it is quite weak while it becomes more prominent in colder stars.
\begin{table*}
\caption{Values of $T_{\rm{eff}}$ and $v\sin i$ used to create the atmospheric
  models and the synthetic profiles for the stars in Tab.~\ref{tab:stelle}. In
  the last two columns are indicated the atmospheric models that best fit the
  observed profiles.} 
\label{tab:parametri} \centering
\begin{tabular}{ccccc}
\hline\hline 
ID   & $T_{\rm{eff}}$ (K) & $v\sin i$ (km/s) & best fit model 4571 & best fit model $b_2$ \\
\hline
HD 73722   & 6400 & 9 & m10\_m60\_100 & m10\_m60\_100 \\
IC 2391 73 & 6200 & 9 & m10\_m60\_100 & m10\_m60\_100 \\
HD 211415  & 5800 & 4 & m10\_m60\_100 & m10\_m60\_100 \\
HD 74497   & 5600 & 7 & m10\_m60\_075 & m10\_m60\_100 \\
IC 2391 98 & 5200 & 13 & m20\_m60\_100 & m10\_m60\_100\\
HD 149661  & 5200 & 3  & m20\_m60\_075 & m10\_m60\_100\\
HD 22049   & 5200 & 8  & m20\_m60\_075 & m10\_m60\_100\\
HD 209100  & 4600 & 3  & m20\_m60\_075 & m10\_m60\_100\\
\hline
\end{tabular}
\end{table*}

As we have explained in Sec.~\ref{sec:atmosphere}, the most active atmospheric
models are the ones with a higher value of $\log m_0$ and/or a steeper
chromospheric temperature gradient. From
Figs.~\ref{fig:star1}-\ref{fig:star8}, we can see that higher activity results
in deeper and broader synthetic profiles. The differences between the
synthetic profiles are more evident in the core of the 4571~{\AA} line that
seems very sensitive to local changes in the atmospheric model, and grow for
lower $T_{\rm{eff}}$. This is in accordance with the results of \citet{langangen}
that found the 4571 {\AA} line a good diagnostic of the lower chromosphere,
especially for cooler atmospheres such as sunspots, where they studied this
line. In particular, in Figs.~\ref{fig:star6} and \ref{fig:star8} we can also
see that a change in the degree of activity can result in a central reversal
of the 4571~{\AA} line. The models often produce a central reversal in this
line that is however not observed and it appears \textbf{when} the formation of the line extends 
above the temperature minimum region, thus depending on the position of the temperature minimum in the models.
\citet{Mg} proposed that this reversal was a
consequence of possible uncertainties in the atomic model whereas \citet{langangen}
and the present work show that increasing the realism in the model atom does
not remove the reversal. Indeed, \citet{langangen} demonstrated that this central
reversal is very sensitive to the position of the chromospheric temperature
rise. Our results show the same dependence of the central reversal from the
the position of the chromospheric temperature rise (higher value of $\log m_0$
corresponding to none or less pronounced reversal), thus confirming those of
\citet{langangen}. Moreover, we find also a dependence of the central reversal
on the chromospheric temperature gradient. A steeper rise of the chromospheric
temperature corresponds to more amount of emission in the line
core. Therefore, on the basis of this result the observations can constrain the structure of the temperature minimum, 
in particular the lowest possible position of the chromospheric temperature rise and its steepest gradient. We note that the 3-D structure of the solar atmosphere could make these constraints less tight. It would be interesting to verify the sensitivity of the 4571 line to the position of the temperature minimum, using solar high spatial and temporal resolution observations of its core but to our knowledge the existing instruments do not have an adeguate spectral resolution to observe core reversals.

We are able to reproduce the \ion{Mg}{i} 4571~{\AA} profile for all the
observed stars with a combination of $T_{\rm{eff}}$, $v\sin i$ and atmospheric
model. Only for two objects, the star HD 73722 and the IC 2391 73
(Fig.~\ref{fig:star1} and \ref{fig:star2}) we cannot reproduce the
observed 4571~{\AA} line, but this could depend from the choice of the
photospheric $T_{\rm{eff}}$ of the star in the atmospheric model, from the value of
$v\sin i$ used for the correction or from the fact that even if we looked for
stars with solar gravity and solar metallicity, these values are an
approximation as we can see from Tab.~\ref{tab:stelle}. Indeed, the star HD 73722
shows the lowest value of $\log g$. Obviously, this comment applies to all the
stars. 

In Sec.~\ref{sec:atomicmodel}, we have discussed the importance of introducing in the calculations some treatment of ultraviolet line-haze opacities. We have shown that without any correction to the {\small{MULTI}} background opacity, the 4571~{\AA} line forms at lower heights in the solar atmosphere becoming less sensitive to changes at the temperature minimum. The effect of this change of formation height, in the case of the stars we are studying, results in small or not significant differences in the synthetic profiles of the 4571~{\AA} line, obtained with different atmospheric models for the same star.

Regarding the $b_2$ profile, there are clear problems in reproducing
the width of the line core. As we have already explained in Sec.~\ref{sec:introduction}, the cores of the $b$
lines form in conditions of NLTE and the populations of the lower and upper levels, involved in their formation, are determined by contributions from photons coming from atmospheric layers other than the ones in which they
form. Therefore, it can be expected that local variations of the atmospheric
structure do not bring large changes in the line intensities.

\section{Conclusions}\label{sec:conclusions}

We analyzed the potential of four \ion{Mg}{i} spectral lines, the
4571~{\AA} line and the $b$ triplet, as diagnostics of chromospheric
activity. Starting from the Next-Gen photospheric models, we built a grid of
atmospheric models and solved the coupled equations of radiative transfer and
statistical equilibrium, obtaining synthetic profiles. These profiles have
been compared with observed spectra of main-sequence, solar like stars with
effective temperatures $T_{\rm{eff}}= 4800, 5200, 5600, 5800, 6200$ and
$6400$~K, solar gravity and solar metallicity.

The comparison between the synthetic spectra and the observations is good for
the \ion{Mg}{i} 4571~{\AA} line for stars with $T<6000$~K while there are
clear problems in reproducing the width of the line core for the $b$
triplet and/or the wings. We believe that this problem could be alleviated, or
even solved, by adopting for each star a photospheric model more closely matching its stellar
parameters (temperature, pressure, gravity and/or metallicity). However, the
grid adopted in this work (Next-Gen) is not sufficiently fine for this purpose
and creating ad hoc photospheric models is outside the scope of this work.

From our analysis, we can conclude that the \ion{Mg}{i} 4571~{\AA} line,
suggested by \citet{langangen} to be a good diagnostic of the lower solar
chromosphere, is significantly sensitive to changes in the activity, obtained
as local changes in the atmospheric model around the minimum of temperature
where the line forms, also in solar-like stars. We find the same dependency of
the amount of emission in the line core of this line from the position of the
chromospheric temperature rise as described by \citet{langangen} and also a
further dependence on the gradient of the chromospheric temperature. A lack of
central reversal in the 4571~{\AA} line gives constraint on the lowest
possible position of the chromospheric temperature rise and its gradient. With this work, we also want to underline the importance of including in the Mg atomic model a correct treatment of ultraviolet line opacities. Without any correction to the {\small{MULTI}} background opacity, the 4571~{\AA} line forms at lower heights and becomes less sensitive to changes in the atmospheric models and, in particular, at the temperature minimum. We verified that, for the stars we have studied in this work, without using any line blanketing treatment, the synthetic profiles of the 4571~{\AA} line obtained with different atmospheric models for the same star, do not differ significantly.

The \ion{Mg}{i} $b$ triplet shows instead very small and not significant
response to the local atmospheric structure, in the same sample of solar-like
stars selected for this work. These lines are not responsive diagnostics for
 atmospheric stratification.

\begin{acknowledgements}
This work was supported by the ASI/INAF contracts I/013/12/0 and I/013/12/1. We thank the referee for useful suggestions and comments.
\end{acknowledgements}

\bibliographystyle{aa}

\begin{thebibliography}{}
\bibitem[Andretta et al. (2005)]{andretta} Andretta, V., Bus\`a, I., Gomez, M. T., \&
  Terranegra, L. 2005, \aap, 430, 669
\bibitem[Anstee \& O'Mara (1995)]{vdw2} Anstee, S. D., \& O'Mara, B. J. 1995,
  \mnras, 276, 859 
\bibitem[Allard \& Hauschildt (1995)]{nextgen} Allard, F., \& Hauschildt,
  P.H. 1995, \apj, 445, 433
\bibitem[Altrock \& Canfield (1974)]{altrock} Altrock, R. C., \& Canfield,
  R. C. 1974, \apj, 194, 733
\bibitem[Altrock \& Cannon (1972)]{cannon} Altrock, R. C., \& Cannon,
  C. J. 1972, \solphys, 26, 21
\bibitem[Athay \& Canfield (1969)]{canfield} Athay, R. G., \& Canfield,
  R. C. 1969, \apj, 156, 695
\bibitem[Athay \& House (1962)]{athay} Athay, R. G., \& House,
  L. L. 1962, \apj, 135, 500
\bibitem[Avrett \& Loeser (2008)]{alc7} Avrett, E. H., \& Loeser,
  R. 2008, \apj, 175, 229	
\bibitem[Bagnulo et al. (2003)]{bagnulo} Bagnulo, S., Jehin, E., Ledoux, C.,
  et. al 2003, The Messanger, 114, 10
\bibitem[Basri et al. (1989)]{basri} Basri, G., Wilcots, E., \& Stout,
  N. 1989, \pasp, 101, 528
\bibitem[Battistini \& Bensby (2015)]{battistini} Battistini, C., \& Bensby,
  T. 2015, \aap, 577, 9
\bibitem[Bus\`a et al. (2001)]{busa} Bus\`a, I., Andretta, V., Gomez, M. T.,
  \& Terranegra, L. 2001, \aap, 373, 993
\bibitem[Bus\`a et al. (2007)]{busa1} Bus\`a, I., Aznar Cuadrado, R.,
  Terranegra, L., Andretta, V., \& Gomez, M. T. 2007,\aap, 466, 1089
\bibitem[Carlsson (1986)]{multi} Carlsson, M. 1986, Uppsala Astron. Obs. Rep.,
  33
\bibitem[Carlsson et al. (1992)]{carlsson} Carlsson, M., Rutten, R. J., \&
  Shchukina, N. G. 1992, \aap, 253, 567
\bibitem[Casagrande et al. (2011)]{casagrande} Casagrande, L., Schoenrich, R.,
  Asplund, M., et al. 2011, \aap, 530, 138
\bibitem[Cenarro et al. (2007)]{cenarro} Cenarro, A. J., Peletier, R. F.,
  Sanchez-Blazquez, P., et al. 2007, \mnras, 374, 664
\bibitem[Deridder \& van Rensbergen (1976)]{vdw1} Deridder, G., \& van
  Rensbergen, W. 1976, \aaps, 23, 147
\bibitem[D'Orazi \& Randich (2009)]{dorazi} D'Orazi, V., \& Randich, S. 2009,
  \aap, 501, 553 
\bibitem[Duncan et al. (1991)]{duncan}Duncan, D. K., Vaughan, A. H., Wilson,
  O. C., et al. 1991, \apjs, 76, 383
 \bibitem[Fontenla et al. (1993)]{falc} Fontenla, J. M., Avrett, E. H., \&
  Loeser, R. 1993, \apj, 406, 319
\bibitem[Glebocki \& Gnaci\`nski (2005)]{glebocki} Glebocki, R., \&
  Gnaci\`nski, P. 2005, 5, VizieR Online Data Catalog: III/244
\bibitem[Gomez da Silva et al. (2014)]{dasilva} Gomez da Silva, J., Santos,
  N. C., Boisse, I., Dumusque, X., \& Lovis, C. 2014, \aap, 566, 66
\bibitem[Hall et al. (2007)]{hall}Hall, J. C., Lockwood, G. W., \& Skiff,
  B. A. 2007, \aj, 133, 862
\bibitem[Heasley \& Allen (1980)]{heasley} Heasley, J. N., \& Allen,
  M. S. 1980, \apj, 237, 255
\bibitem[Henry et al. (1996)]{henry} Henry T. J., Soderblom D. R., Donahue,
  R. A., \& Baliunas, S. L. 1996, \aj, 111, 439 
\bibitem[Katsova \& Livshits (2011)]{katsova} Katsova, M. M., \& Livshits,
  M. A. 2011, Astronomy Reports, Vol. 55, Issue 12, 1123
\bibitem[Kurucz et al. (1984)]{atlas} Kurucz, R. L., Furenlid, I., Brault, J.,
  \& Testerman, L. 1984, Solar flux atlas from 296 to 1300 nm (National Solar
  Observatory)
\bibitem[Langangen et al. (2009)]{langangen} Langangen, {\O}., \& Carlsson,
  M. 2009, \apj, 696, 1892
\bibitem[Liu et al. (2015)]{liu} Liu, C., Ruchti, G., Feltzing, S, et
  al. 2015, \aap, 575, A51
\bibitem[Magain (1986)]{magain} Magain, P.  1986, \aap, 163, 135
\bibitem[Marsden et al. (2009)]{marsdena} Marsden, S. C., Carter, B., D., \&
  Donati,  J. F. 2009, \mnras, 339, 888
\bibitem[Marsden et al. (2014)]{marsdenb} Marsden, S. C., Petit, P., Jeffers,
  S. V., et al. 2014, \mnras, 444, 3517
\bibitem[Mashonkina et al. (1996)]{mashonkina} Mashonkina, L. I., Shimanskaya,
  N. N., \&  Sakhibullin, N. A. 1996, A. Rep., 40, 187
\bibitem[Mauas et al. (1988)]{Mg} Mauas, P. J., Avrett, E. H., \& Loeser,
  R. 1988, \apj, 330, 1008 
\bibitem[Mermilliod et al. (2009)]{mermilliod} Mermilliod, J. C., Mayor, M.,
  \& Udry, S. 2009, \aap, 498, 949
\bibitem[Merle et al. (2011)]{merle1} Merle, T., Thévenin, F., Pichon, B., \&
  Bigot, L. 2011, \mnras, 418, 863
\bibitem[Merle et al. (2013a)]{merle2} Merle, T., Thévenin, F., Belyaev,
  A. K. et al. 2013, in New Advances in Stellar Physics: From Microscopic to
  Macroscopic Processes, eds.: G. Alecian, Y. Lebreton, O. Richard, and
  G. Vauclair, EAS Publications Series, vol. 63, 331
\bibitem[Merle et al. (2013b)]{merle3} Merle, T., Thévenin, F., Guitou, M., et
  al. 2013, in SF2A-2013: Proceedings of the Annual meeting of the French
  Society of Astronomy and Astrophysics, eds.: L. Cambresy, F. Martins,
  E. Nuss, A. Palacios, 247
\bibitem[Milahas (1978)]{mihalas} Mihalas, D. 1978, Stellar Atmospheres, 2nd
  edn. (San Francisco: Freeman \& Co)
\bibitem[Montes et al. (1999a)]{monteslibrary} Montes, D., Ramsey, L. W., \&
  Welty, A. D. 1999, \apjs, 123, 283 
\bibitem[Montes et al. (1999b)]{montes} Montes, D., Saar, S. H., Collier
  Cameron, A., \& Unruh, Y. C. 1999, \mnras, 305, 45
\bibitem[Moore et al. (1966)]{moore} Moore, C. E., Minnaert, M. G. J., \& Houtgast, J. 1966, The solar spectrum 2935 A to 8770~{\AA}, National Bureau of Standards Monograph, Washington: US Government Printing Office
\bibitem[N\"ordstrom et al. (2004)]{nordstrom} N\"ordstrom, B., Mayor, M.,
 Andersen, J., et al. 2004, \aap, 418, 989
\bibitem[Noyes et al. (1984)]{noyes}Noyes, R. W., Hartmann, L. W., Baliunas,
  S. L., Duncan, D. K., \& Vaughan, A. H. 1984, \apj, 279, 763
\bibitem[Osorio et al. (2015)]{osorio1} Osorio, Y., Barklem, P. S., Lind, K.,
  et al. 2015, 579, A53
\bibitem[Osorio \& Barklem (2016)]{osorio2} Osorio, Y., \& Barklem,
  P. S. 2016, 586, A120 
\bibitem[Sasso (2004)]{sasso} Sasso, C. 2004, Thesis
\bibitem[Shimanskaya et al. (2000)]{shimanskaya} Shimanskaya, N. N.,
  Mashonkina, L. I., \& Sakhibullin, N. A. 2000, A. Rep., 44, 530
\bibitem[Soderblom et al. (1993)]{soderblom} Soderblom, D. R., Stauffer, J. R.,
  Hudon, J. D., \& Jones, B. F. 1993, \apjs, 85, 315
\bibitem[Soubiran et al. (2008)]{soubiran} Soubiran, C., Bienaym\'e, O.,
  Mishenina, T. V., \& Kovtyukh, V. V. 2008, \aap, 480, 91
\bibitem[Stauffer et al. (1989)]{stauffera} Stauffer, J. R., Hartmann, L. W.,
  Prosser, C. F., et al. 1997, \apj, 479, 776
\bibitem[Vernazza et al. (1981)]{VAL3C} Vernazza, J. E., Avrett, E. H., \&
  Loeser, R. 1981, \apjs, 45, 635 
\bibitem[White et al. (1972)]{white} White, O. R., Altrock, R. C., Brault,
  J. W., \& Slaughter, C. D. 1972, \solphys, 23, 18
\bibitem[Wilson (1978)]{wilson}Wilson, O. C. 1978, \apj, 226, 379
\bibitem[Zhao \& Gehren (2000)]{zhao1} Zhao, G., \& Gehren, T. 2000, \aap,
  362, 1077 
\bibitem[Zhao et al. (1998)]{zhao2} Zhao, G., Butler, K., \& Gehren, T. 1998,
  \aap, 333, 219 
\end{thebibliography}

\end{document}